\documentclass[aps,prl,twocolumn,superscriptaddress,showpacs,,reprint]{revtex4-1}
\usepackage{graphicx}
\usepackage{latexsym}    
\usepackage{amssymb} 
\usepackage{mathrsfs}
\usepackage{amsmath}
\usepackage{subfigure}
\usepackage{bm}
\usepackage{verbatim}
\usepackage{color}
\usepackage{xcolor}
\usepackage{hyperref}
\usepackage{epsfig}
\usepackage{multirow}

\begin{document}

\title{Topological Phonon Modes in A Two-Dimensional Wigner Crystal}

\author{Wencheng Ji}

\affiliation{International Center for Quantum Materials, Peking University, Beijing
100871, China}

\author{Junren Shi}
\email{junrenshi@pku.edu.cn}

\affiliation{International Center for Quantum Materials, Peking University, Beijing
100871, China}

\affiliation{Collaborative Innovation Center of Quantum Matter, Beijing 100871,
China}
\begin{abstract}
We investigate the spin-orbit coupling effect in a two-dimensional
Wigner crystal. We show that sufficiently strong spin-orbit coupling
and an appropriate sign of $g$-factor could transform the Wigner
crystal to a topological phonon system. We demonstrate the existence
of chiral phonon edge modes in finite size samples, as well as the
robustness of the modes in the topological phase. We explore the possibility
of realizing the topological phonon system in two-dimensional Wigner
crystals confined in semiconductor quantum wells/heterostructure.
We find that the spin-orbit coupling is too weak for driving a topological
phase transition in these systems. We argue that one may look for
the topological phonon system in correlated Wigner crystals with emergent
effective spin-orbit coupling. 
\end{abstract}

\maketitle
The concept of topology attracts great interest in recent investigations
of condensed matter physics. The quantum Hall effect demonstrates
a new state of matter which carries robust chiral edge modes protected
by non-trivial topology of electron states. Recent research reveals
various kind of topological insulators dictated by electron state
topology in the presence of the time reversal symmetry or other symmetries~\cite{key-1a,key-1b,key-1c,key-1d}.
Up to now, the studies are mainly focused on topological effects of
the charge degree of freedom of electrons. On the other hand, the
interest is recently extended to topological effects associated with
collective excitations such as phonons and magnons~\cite{key-2a,key-2b,key-2c,key-2d,Strohm2005,Inyushkin2007}.
Particularly, a topological phonon system (TPS) is predicted to possess
robust chiral vibrational (phonon) modes at its edge~\cite{key-2a}.
These topological phonon modes (TPMs) are believed to be contributing
to quantized thermal hall conductivity at low temperature~\cite{key-2d}.
It is even postulated that TPMs are ubiquitous in biological systems
and essential for a variety of processes in living organisms~\cite{key-2a}.
While these theoretical considerations are interesting and intriguing,
we have yet to know an experimentally accessible and controllable
way to construct a TPS in laboratory.

In this paper, we explore the possibility of realizing a TPS in a
Wigner crystal (WC) of electrons. We investigate the effect of spin-orbit
coupling (SOC) in a two-dimensional WC~\cite{Monarkha2012}. We show
that sufficiently strong SOC and an appropriate sign of $g$-factor
could transform the WC to a TPS. We demonstrate the existence of chiral
edge phonon modes in finite size samples, as well as the robustness
of the chiral edge modes in the topological phase. We check the possibility
in two-dimensional WCs confined in semiconductor quantum wells/heterostructure~\cite{Ando1982}.
We find that the SOC is too weak for driving a topological phase transition
in these systems. We argue that one may look for the TPS in correlated
WCs with emergent effective SOC. 

It is well known that electrons form a WC in sufficiently low density
when the Coulomb interaction dominates over the kinetic energy. Recent
state-of-art Monte Carlo calculations show that two-dimensional electrons
undergo a series of phase transitions when lowering density, ultimately
stabilized to a WC phase of a triangular lattice with ferromagnetic
order of electron spins when $r_{s}>38$~\cite{Tanatar1989,key-7-2,key-7-3},
where $r_{s}\equiv1/\sqrt{\pi n_{e}}a_{B}$ is a dimensionless density
parameter with electron density $n_{e}$ and effective Bohr radius
$a_{B}$. Experimentally, two-dimensional WCs had been observed in
ultra-clean 2D samples of GaAs/AlGaAs heterostructure~\cite{Yoon1999}.

We investigate phonon modes of a 2D WC of electrons. To model such
a system, we consider a set of electrons with an effective mass $m^{\ast}$
arranged in a triangular lattice vibrating near their equilibrium
positions. The Hamiltonian can be written as: 
\begin{align}
H= & \sum_{\bm{l}}\frac{\hat{\bm{P}}(\bm{l})^{2}}{2m^{\ast}}+\Phi[\{u_{\alpha}(\bm{l})\}]+\sum_{\bm{l}}g^{\ast}\mu_{B}\hat{\sigma}_{\bm{l}z}B\,,\label{hamiltonian}
\end{align}
where the first term is the kinetic energy, the second term is the
potential energy due to the Coulomb interaction between electrons,
and the third term is the Zeeman energy due to a uniform magnetic
field $B$ perpendicular to the 2D plane, with $\mu_{B}$ being Bohr
magneton, $g^{\ast}$ being effective Landé factor for the specific
material hosting the 2D system, and $\hat{\sigma}_{\bm{l}z}$ being
the $z$-component of the Pauli spin matrices $\hat{\bm{\sigma}}_{\bm{l}}$
for an electron at site $\bm{l}$. For an electron subjected to both
the magnetic field $\bm{B}$ and SOC, the kinetic momentum should
be written as, 
\begin{equation}
\hat{\bm{P}}(\bm{l})=-i\hbar\bm{\nabla}_{\bm{u}(\bm{l})}+\frac{1}{2}u(\bm{l})\times e\bm{B}+\alpha_{0}\hat{\bm{\sigma}}_{\bm{l}}\times\bm{F}(\bm{l}),\label{Pl}
\end{equation}
which involves two extra terms besides the canonical momentum operator,
consisting of a vector potential term induced by the perpendicular
magnetic field $\bm{B}$ and a SOC term induced by the electrostatic
force $\bm{F}(\bm{l})$ acting on the electron by all other electrons
in the system, respectively, and $u(\bm{l})$ is the displacement
of an electron relative to its equilibrium position. The strength
of the SOC is determined by a coefficient $\alpha_{0}$, which is
a material-specific parameter~\cite{key-6b}. Because all electrons
are localized near their equilibrium positions in a WC, it is sufficient
to make a harmonic approximation to the potential energy: 
\begin{equation}
\Phi[\{u_{\alpha}(\bm{l})\}]\approx\Phi_{0}+\frac{1}{2}\sum_{\bm{l}\bm{l}',\alpha,\beta}\Phi_{\alpha\beta}^{\bm{l}-\bm{l}'}u_{\alpha}(\bm{l})u_{\beta}(\bm{l}')\,,\label{potential}
\end{equation}
where $\ensuremath{\alpha,\beta}=x,y$, and the detailed form of the
dynamic coefficients $\Phi_{\alpha\beta}^{\bm{l}-\bm{l}'}$ for a
triangular lattice is presented in Ref.~\cite{key-5a}. Correspondingly,
the force $\bm{F}(\bm{l})$, which dictates the SOC, is related to
the potential energy by: 
\begin{equation}
F_{\alpha}(\bm{l})=-\nabla_{u_{\alpha}(\bm{l})}\Phi=-\sum_{\bm{l}',\beta}\Phi_{\alpha\beta}^{\bm{l}-\bm{l}'}u_{\beta}(\bm{l}').
\end{equation}

To proceed, we need to make a further approximation. Equation (\ref{hamiltonian})
defines a system involving both phonons (vibration modes) and magnons
(spin waves). However, since the WC has ferromagnetic order of spins
and a perpendicular magnetic field will align and quench the spin
degree of freedom, we can replace the Pauli spin matrices in the Hamiltonian
with their mean field expectation values: $\left\langle \hat{\bm{\sigma}}_{\bm{l}}\right\rangle =\left(0,0,\sigma\right)$
with $\sigma=-\mathrm{sgn}(g^{\ast}B)$, because the direction of
the electron spins will be parallel (anti-parallel) to the external
magnetic field for $g^{\ast}<0$ ($g^{\ast}>0$).

We can then obtain equations of motion:
\begin{align}
\dot{u}_{\alpha}(\bm{l})= & \frac{\hat{P}_{\alpha}(\bm{l})}{m^{\ast}},\\
\dot{\hat{P}}_{\alpha}(\bm{l})= & -\sum_{l'\beta}\left[\Phi_{\alpha\beta}^{\bm{l}-\bm{l}'}u_{\beta}(\bm{l}')+G_{\bm{l}-\bm{l}^{\prime}}\epsilon_{\alpha\beta}\mathit{\hat{P}_{\beta}}(\bm{l}')\right],
\end{align}
where $\epsilon_{\alpha\beta}$ is antisymmetric tensor with $\epsilon_{xy}=1$,
and 
\begin{align}
G_{\bm{l}-\bm{l}^{\prime}}= & \omega_{c}\delta_{\bm{l},\bm{l}'}-\frac{\alpha_{0}\sigma}{m^{\ast}}\left[\Phi_{yy}^{\bm{l}-\bm{l}^{\prime}}+\Phi_{xx}^{\bm{l}-\bm{l}^{\prime}}\right]
\end{align}
where $\omega_{c}=eB/m^{\ast}$ is the magnetic cyclotron frequency
of electrons.

A Fourier transformation recasts the equations of motion to an eigenvalue
equation for phonon modes in a system breaking time-reversal symmetry~\cite{key-2d}:
\begin{equation}
\left[\begin{array}{cc}
0 & i\\
-i\Phi(\bm{k}) & G(\bm{k})\sigma_{2}
\end{array}\right]\psi_{n}\left(\bm{k}\right)=\omega_{n}\left(\bm{k}\right)\psi_{n}\left(\bm{k}\right),\label{eigeneq}
\end{equation}
where $\bm{k}$ is defined in the Brillouin zone for the triangular
lattice, $\psi_{n}\left(\mathbf{k}\right)=\left(\bm{u}(\bm{k}),\bm{P}(\bm{k})\right)^{T}$,
$\sigma_{2}$ is the second Pauli matrix, and $\Phi(\bm{k})$, $G(\bm{k})$,
$\bm{u}(\bm{k})$, and $\bm{P}(\bm{k})$ are Fourier transformations
of $\Phi(\bm{l}-\bm{l}')/{m^{\ast}}$, $G_{\bm{l}-\bm{l}^{\prime}}$,
${m^{\ast}}\bm{u}(\bm{l})$, and $\bm{P}(\bm{l})$, respectively.
We have:
\begin{equation}
G(\bm{k})=\omega_{c}-\alpha_{0}\left(\Phi_{xx}(\bm{k})+\Phi_{yy}(\bm{k})\right)\sigma.
\end{equation}
We note that Eq.~(\ref{eigeneq}) will give rise to four modes for
each $\bm{k}$ with two positive and two negative frequency branches.
However, the negative frequency branches can be related to the positive
frequency branches by $\omega_{n}^{-}(\bm{k})=-\omega_{n}^{+}(-\bm{k})$~\cite{key-2d}.
In the following, we will only show the positive frequency branches.

\begin{figure}[h]
\includegraphics[width=0.5\columnwidth]{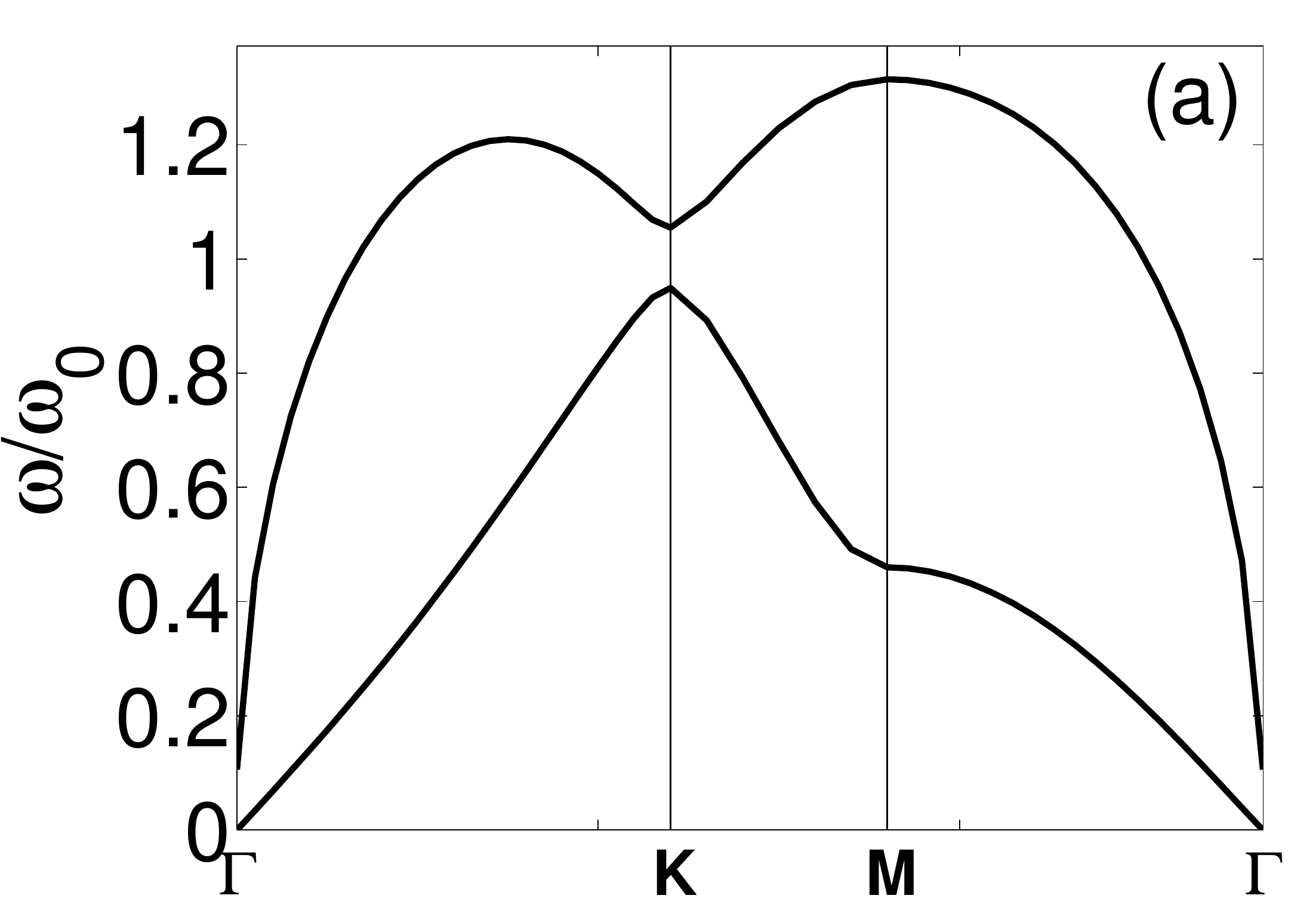}\includegraphics[width=0.5\columnwidth]{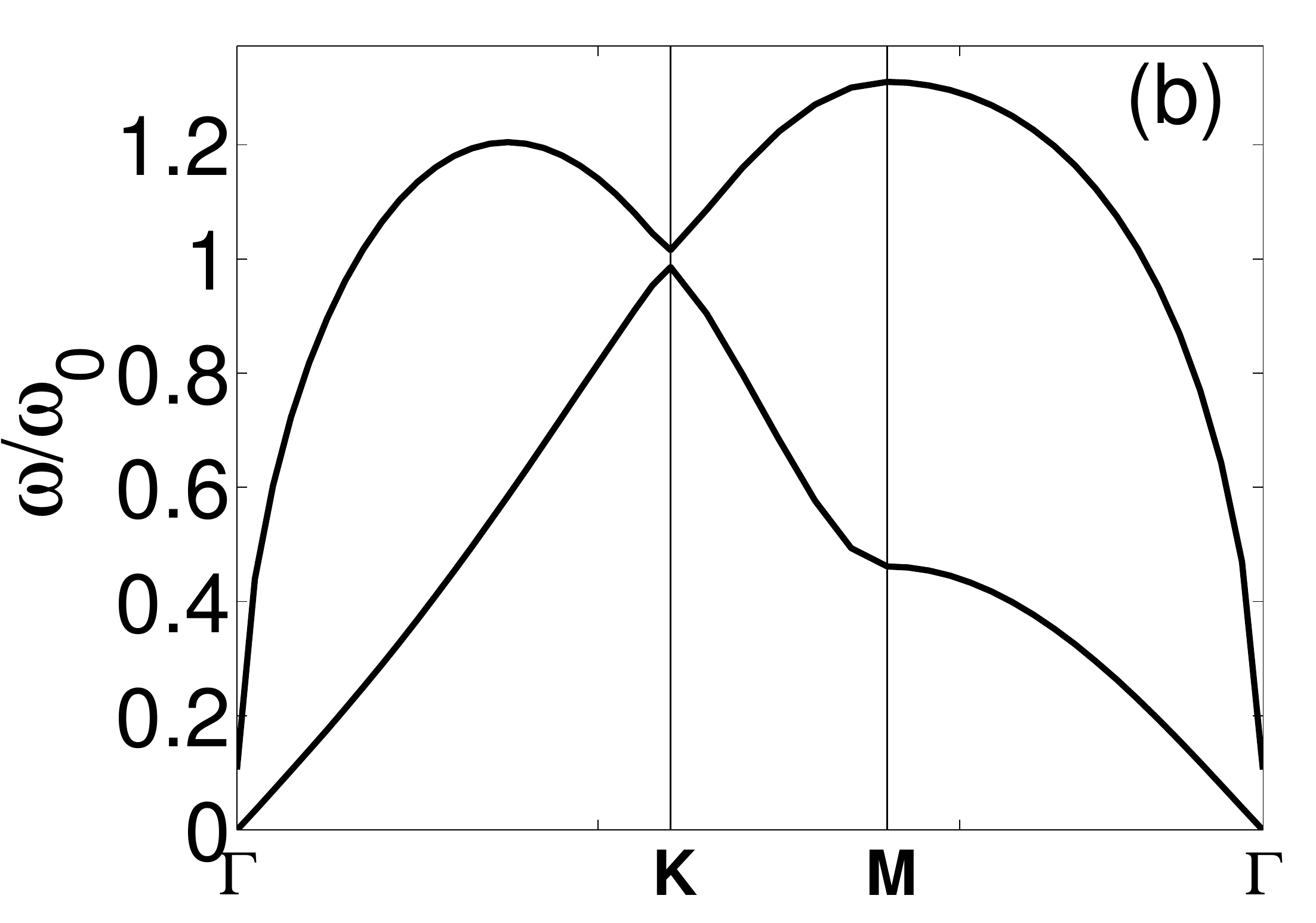}

\includegraphics[width=0.5\columnwidth]{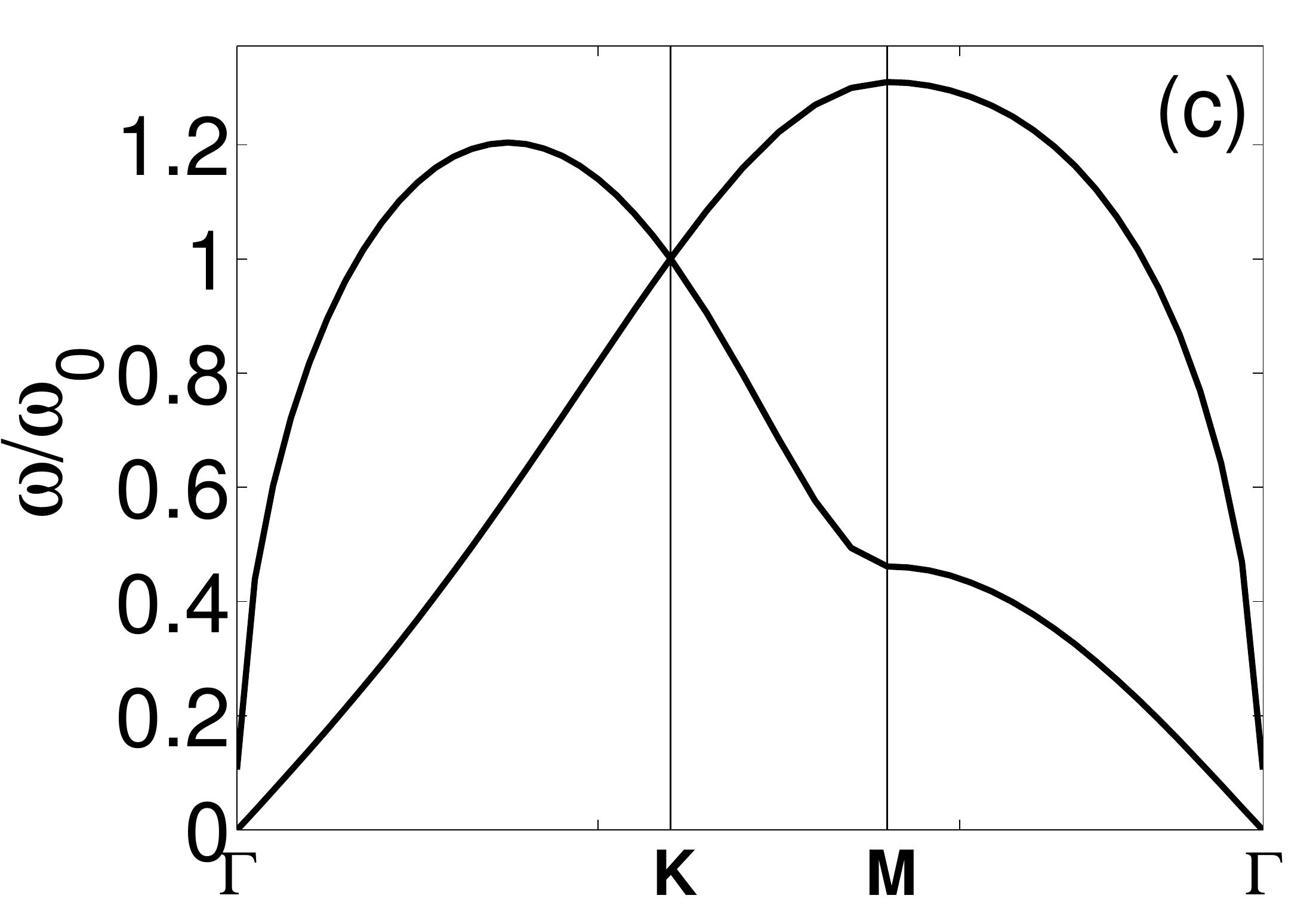}\includegraphics[width=0.5\columnwidth]{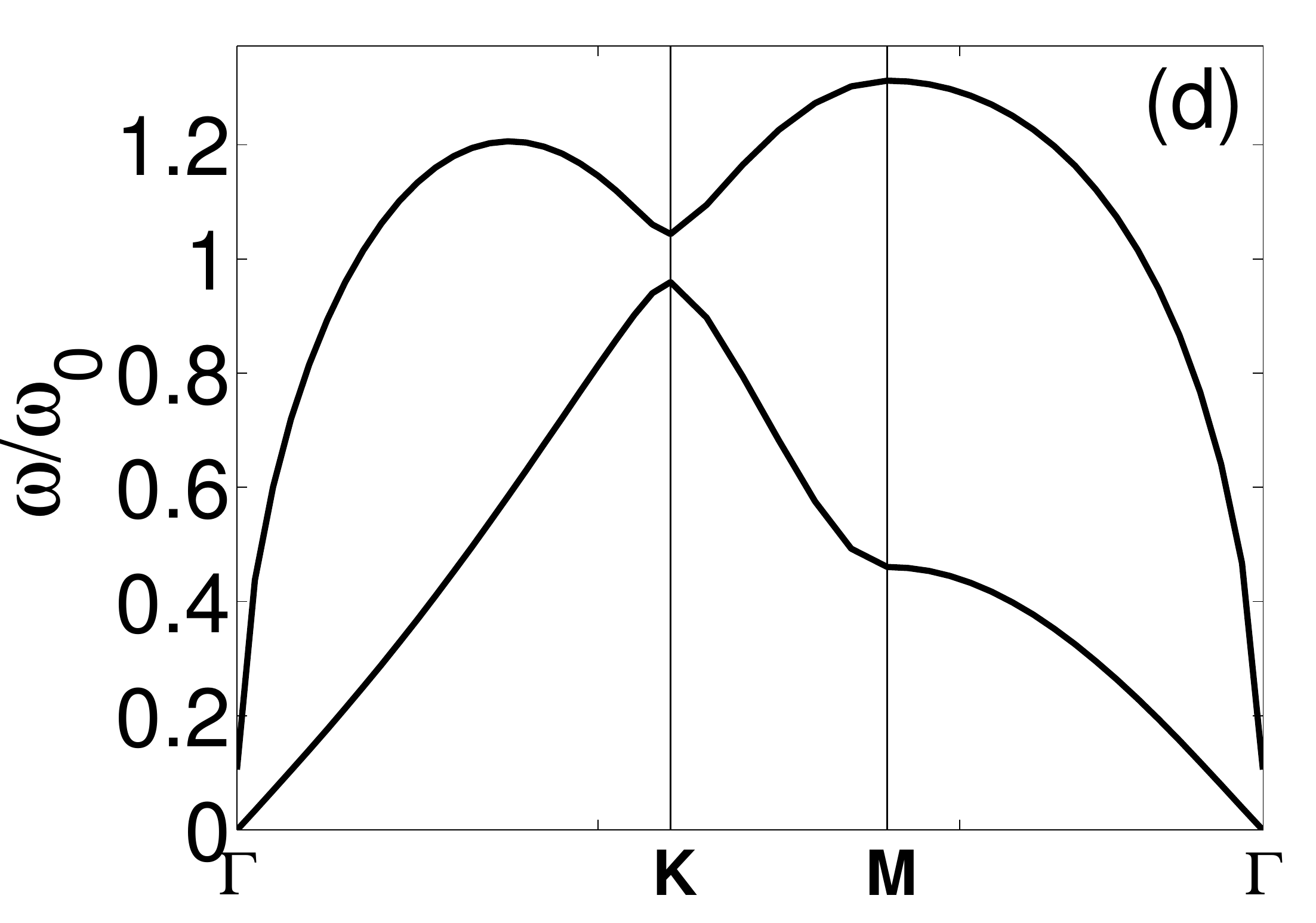}

\caption{Phonon dispersion along high symmetry lines of the Brillouin zone
for different strengths of SOC. We set $\omega_{c}=0.1\omega_{0}$.
The strengths of SOC are: (a) $\alpha_{0}=0$; (b) $\alpha_{0}\omega_{0}=0.04$;
(c) $\alpha_{0}\omega_{0}=0.05$; (d) $\alpha_{0}\omega_{0}=0.1$.
The effective Landé factor is assumed to be negative: $g<0$.}
\label{dispersion} 
\end{figure}

Figure \ref{dispersion} shows the evolution of phonon dispersion
along high symmetry lines of the Brillouin zone for different strengths
of SOC in the presence of a magnetic field. We observe that the uniform
magnetic field induces gaps between the two branches of phonon modes
at both $\Gamma$ and $\bm{K}$ ($\bm{K}^{\prime}$) points of the
Brillouin zone, as shown in Fig.~\ref{dispersion}(a). In increasing
the strength of SOC, the gap at $\bm{K}$ ($\bm{K}^{\prime}$) point
is gradually closed and reopened, while the gap at $\Gamma$-point
is unaffected, as shown in Fig.~\ref{dispersion}(b)\textendash (d).
Such behavior is similar to what happens in a topological insulator,
in which a closing and a reopening of a gap indicates a band inversion
which transforms a normal insulator to a topological insulator. We
thus expect that the evolution observed here may also drive a topological
phase transition, albeit for the phonon bands.

\begin{figure}[t]
\includegraphics[width=0.5\columnwidth]{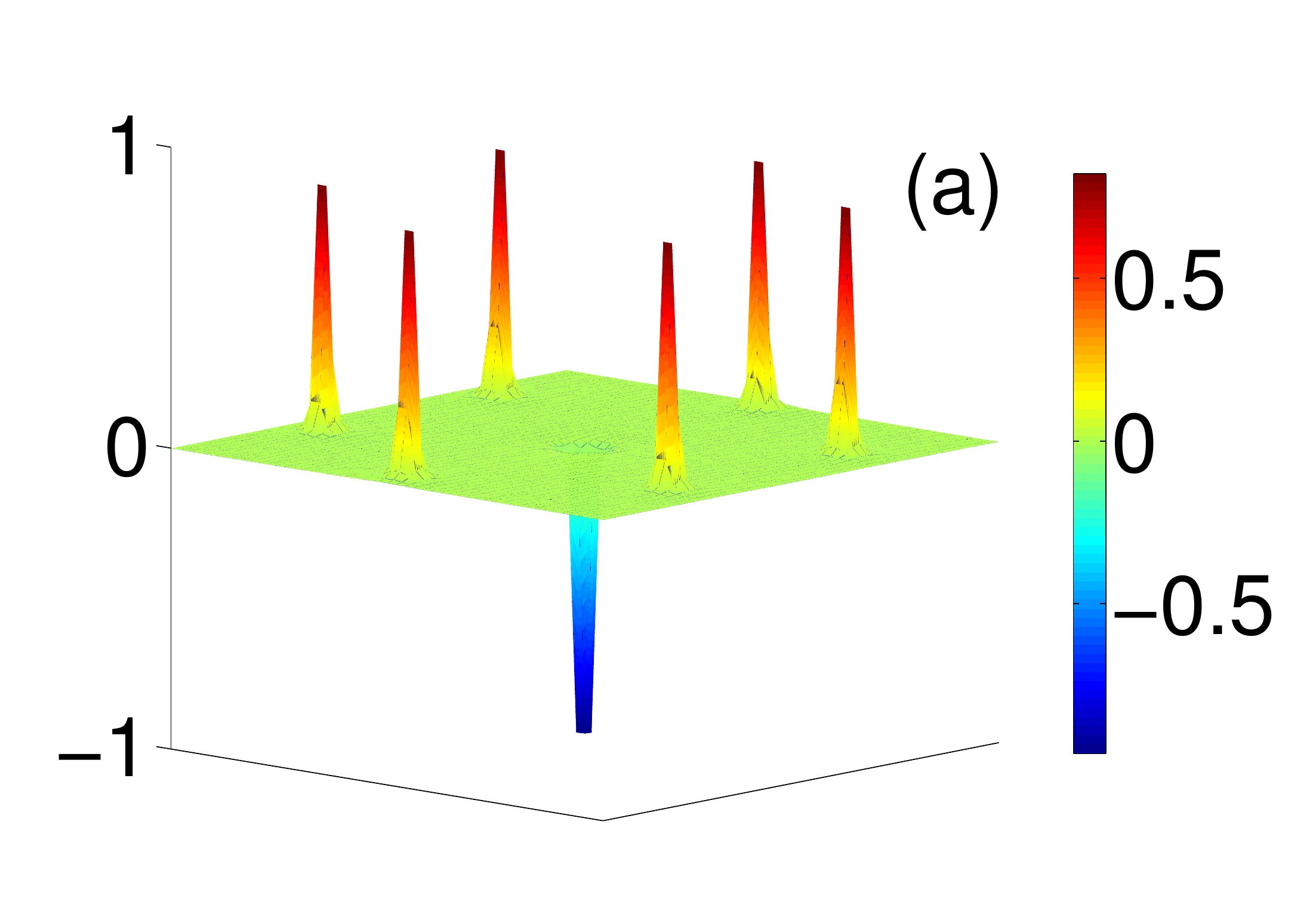}\includegraphics[width=0.5\columnwidth]{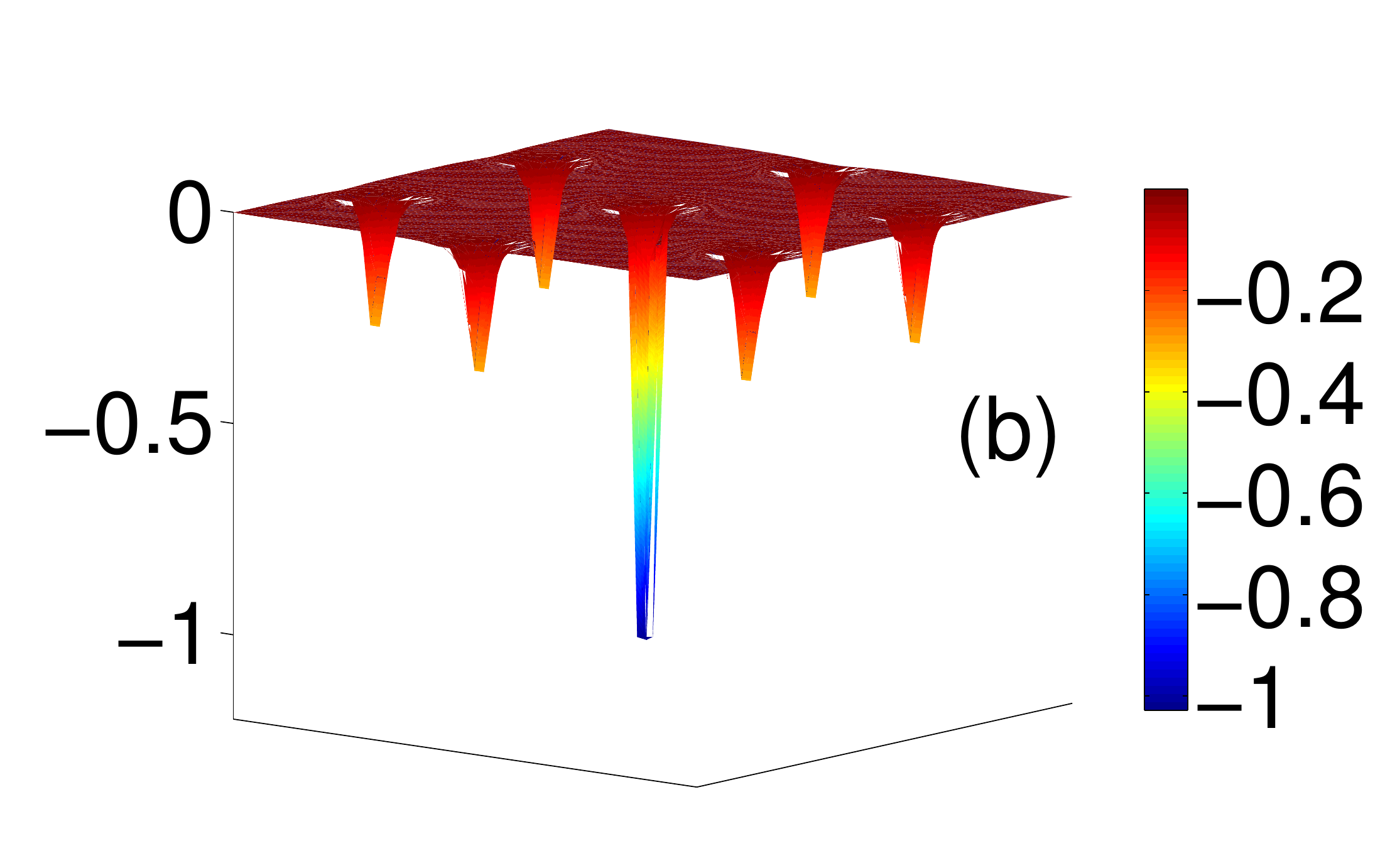}\caption{ (color online) Distribution of the Berry curvatures for the upper
phonon bands before and after the band inversion. The parameters are
the same as those in Fig.~\ref{dispersion}(b) and Fig.~\ref{dispersion}(d),
respectively. }
\label{berry} 
\end{figure}

To confirm the topological change of the phonon bands, we calculate
the phonon Berry curvature that is defined as: 
\begin{align}
\bm{\Omega}_{n}(\bm{k}) & =-\mathrm{Im}\,\left[\frac{\partial\bar{\psi}_{n}\left(\bm{k}\right)}{\partial\bm{k}}\times\frac{\partial\psi_{n}\left(\bm{k}\right)}{\partial\bm{k}}\right],
\end{align}
where $\bar{\psi}_{n}\left(\bm{k}\right)\equiv\psi_{n}^{\dagger}\left(\bm{k}\right)\left[\begin{array}{cc}
\Phi(\bm{k}) & 0\\
0 & I_{2\times2}
\end{array}\right]$ \cite{key-2d}. Figure \ref{berry} shows the distribution of the
Berry curvatures for the upper phonon bands before and after the band
inversion. It can be clearly seen that the Berry curvature peaks at
$\Gamma$ and $\bm{K}$ ($\bm{K}^{\prime}$). For weak SOC, the Berry
curvatures at $\Gamma$ and $\bm{K}$ ($\bm{K}^{\prime}$) have opposite
signs, resulting in cancellation of the Chern number that is proportional
to an integration of the Berry curvature over the whole Brillouin
zone. On the other hand, the Berry curvatures at $\Gamma$ and $\bm{K}$
($\bm{K}^{\prime}$) have the same signs after the band inversion,
indicating the topological change of the phonon band. We calculate
the Chern numbers for both cases, and obtain $C=0$ and $C=-2$ for
bands before and after the inversion, respectively.

\begin{figure}[tbh]
\includegraphics[width=0.6\columnwidth,height=4cm]{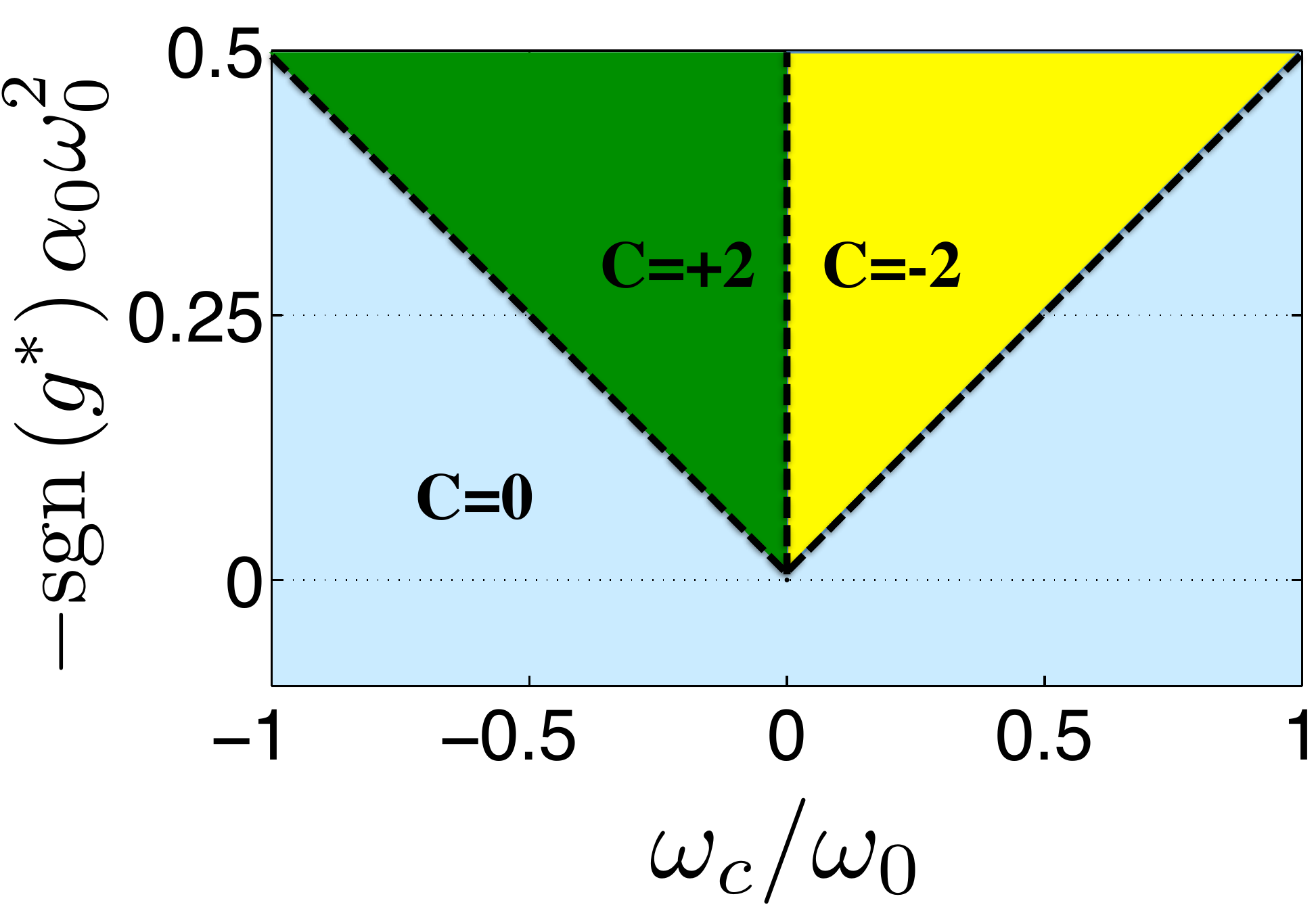} \caption{ \label{pd}(color online) Phase diagram for the phonon bands of a
two-dimensional ferromagnetic WC. The regions with non-zero Chern-number
is topologically nontrivial. }
\end{figure}

The behavior can be easily understood. To see that, we recast the
eigenvalue equation (\ref{eigeneq}) to an equivalent $2\times2$
hermitian eigenvalue problem: 
\begin{equation}
\hat{h}(\bm{k})\mathbf{u}(\bm{k})\equiv\left(\Phi(\bm{k})+\omega G(\bm{k})\sigma_{2}\right)\mathbf{u}(\bm{k})=\omega^{2}\mathbf{u}(\bm{k}).\label{eq:22}
\end{equation}
Near the $\bm{K}$ point, we can expand $\hat{h}(\bm{k})\approx h_{0}+\lambda_{1}\triangle k_{y}\sigma_{1}-\lambda_{2}\triangle k_{x}\sigma_{3}+M\sigma_{2}$,
where $h_{0}$, $\lambda_{1}$, $\lambda_{2}$ are constants, and
\begin{equation}
M=\omega G(\bm{K})=\omega\left(\omega_{c}-2\alpha_{0}\sigma\omega_{0}^{2}\right),
\end{equation}
where $\omega_{0}=\sqrt{\Phi_{xx}(\bm{K})}$ is the phonon frequency
at $\bm{K}$ point in the absence of both the magnetic field and SOC,
and is related to material parameters by $\hbar\omega_{0}=1.967r_{s}^{-3/2}\mathrm{Ry}^{\ast}$,
where $\mathrm{Ry}^{\ast}$ is the effective Rydberg of the hosting
semiconductor~\cite{key-6b}. It is easy to see that the effective
hamiltonian has the same form as the 2D massive Dirac hamiltonian
with the mass $M$. The inversion of the bands and the topological
phase transition occur when $M$ changes sign. This is possible only
when $g^{\ast}\alpha_{0}<0$. In this case, $M$ changes sign when
\begin{equation}
2\alpha_{0}\omega_{0}=-\mathrm{sgn}(g^{\ast})\frac{\omega_{c}}{\omega_{0}}.\label{alphac}
\end{equation}
On the other hand, the Berry curvature near the $\Gamma$ point is
only determined by the external magnetic field and is not affected
by the SOC, because $\Phi_{\alpha\beta}(\bm{\Gamma})=0$ and $G(\bm{\Gamma})=\omega_{c}$.
Figure~\ref{pd} shows the corresponding phase diagram. 

\begin{figure}[h]
\includegraphics[width=0.5\columnwidth]{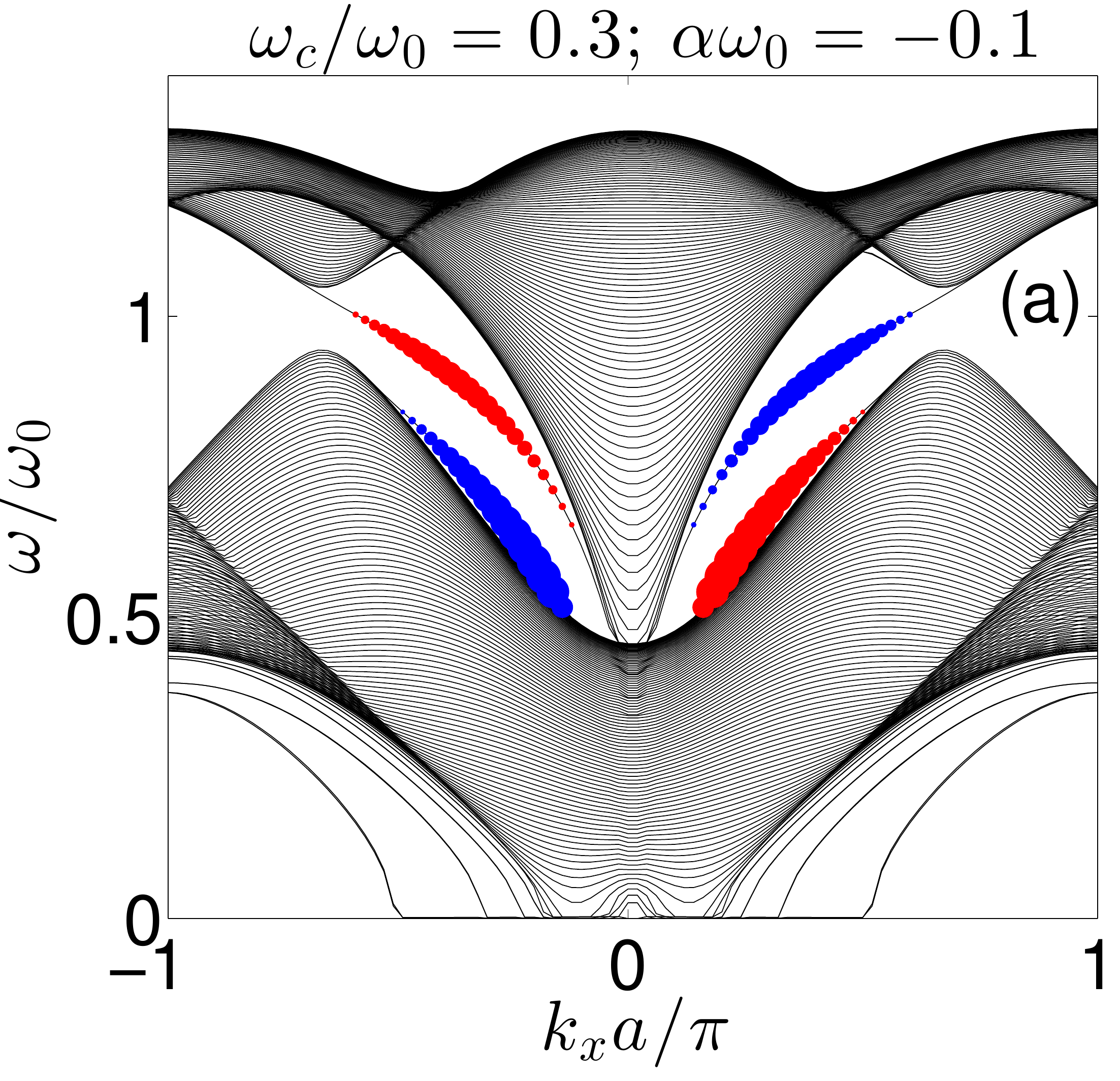}\includegraphics[width=0.5\columnwidth]{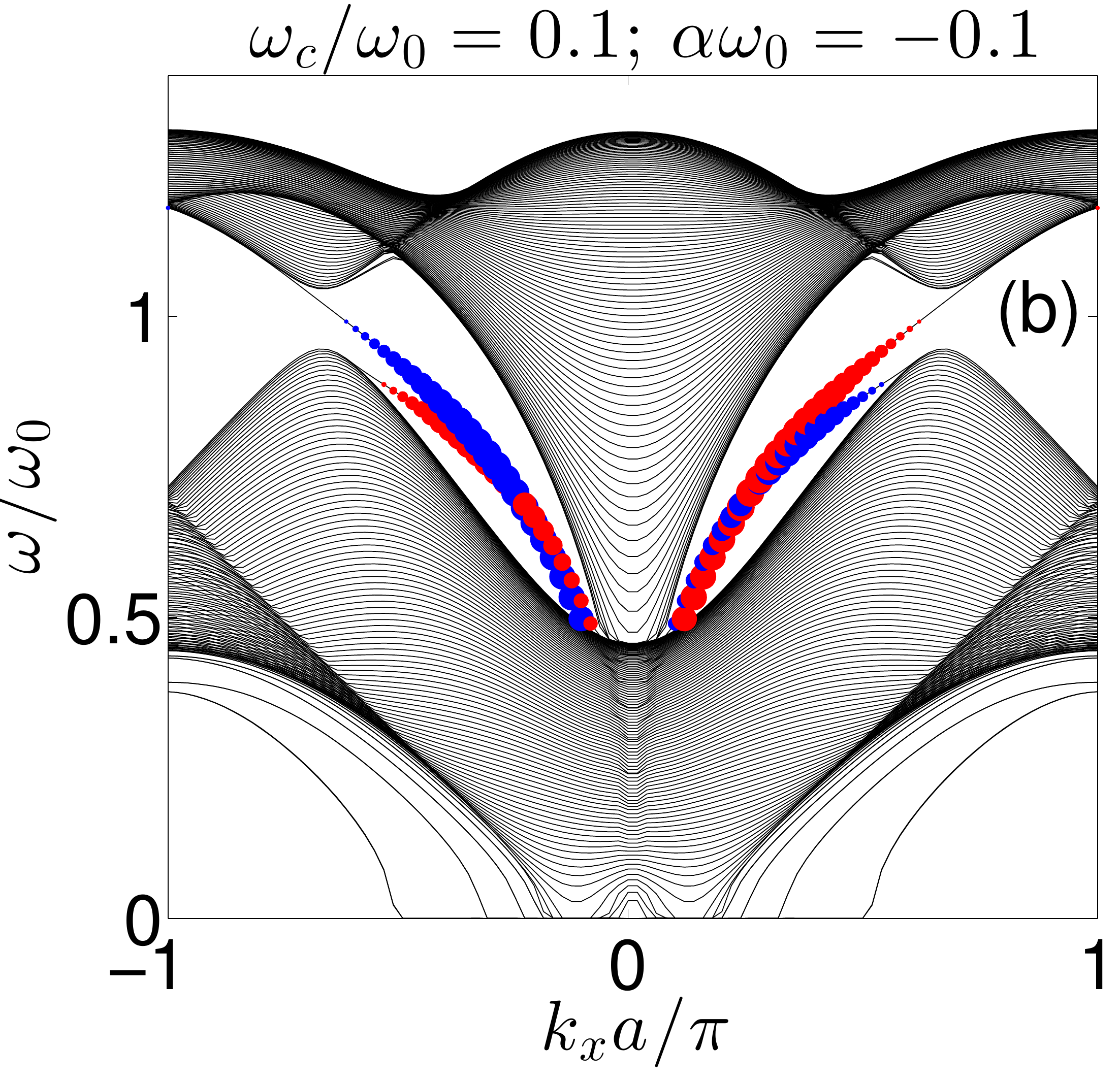}
\includegraphics[width=0.5\columnwidth]{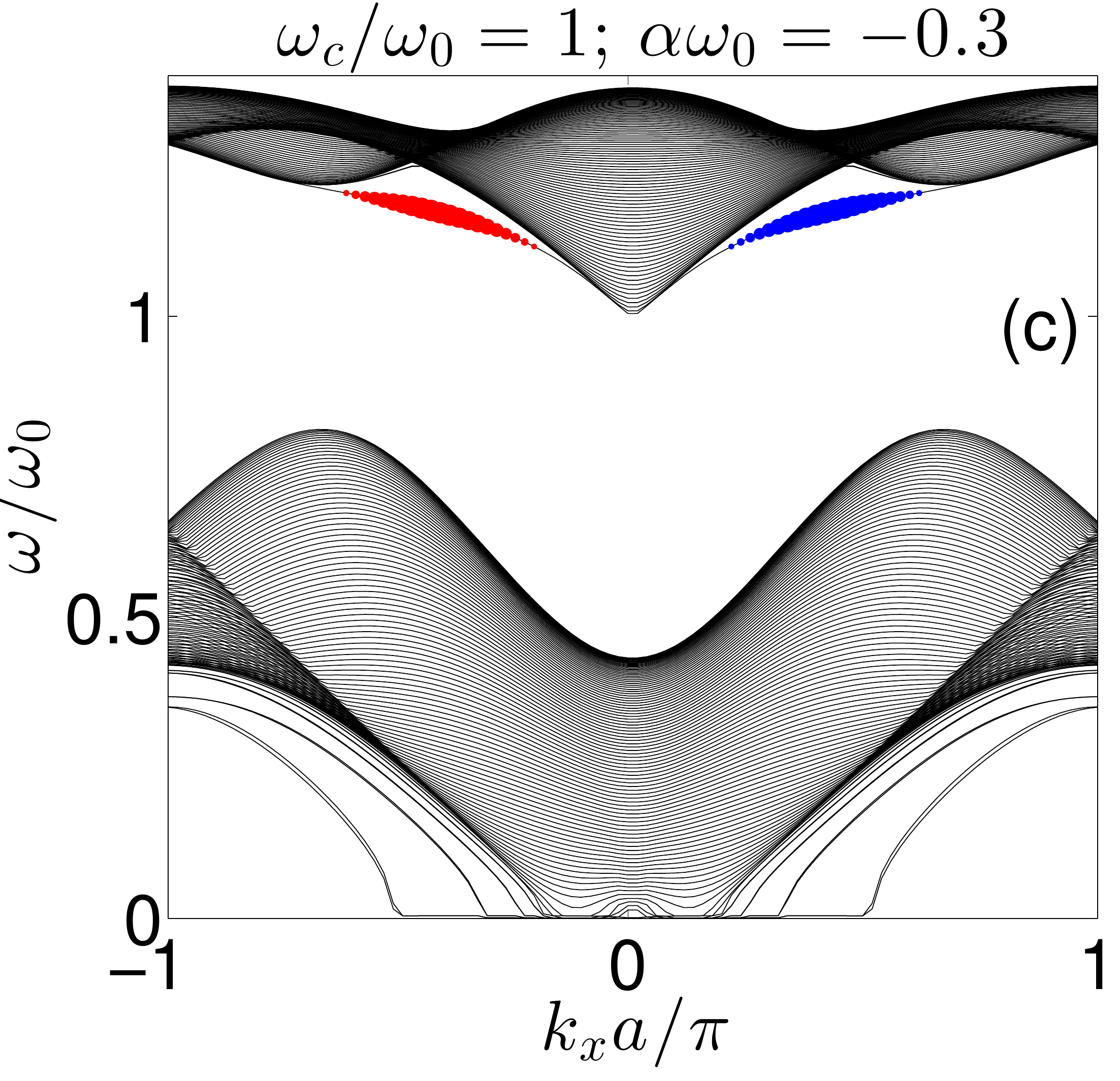}\includegraphics[width=0.5\columnwidth]{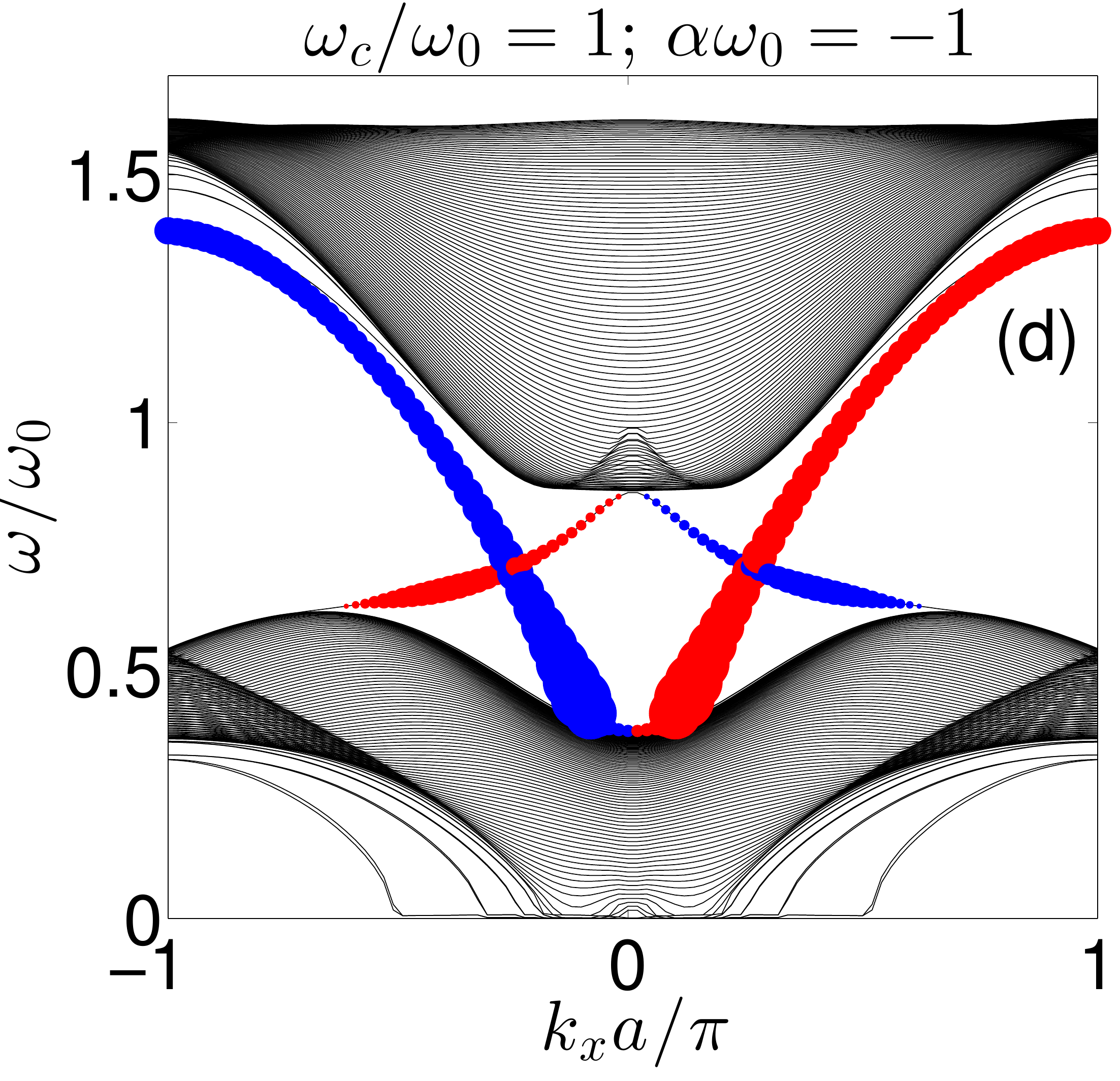}
\caption{\label{Finite}(color online) Phonon dispersion of a finite size sample
with different values of $\alpha\omega_{0}$ and $\omega_{c}/\omega_{0}$.
The finite size sample is extended along the $x$ direction, and has
a width of $100$ sites in the $y$-direction. The red (blue) dots
represent modes localized at the top (bottom) edge of the sample.
The size of a dot indicates the degree of localization that is proportional
to a quantity $e_{n}=\sum_{i}\left|\psi_{n}(i)\right|^{4}$, a larger
value of which corresponds to a more localized edge mode. }
\end{figure}

For a TPS, there exist TPMs in a finite-size sample. To show that,
we consider a strip of WC along the $x$ direction and calculate phonon
dispersion as a function of $k_{x}$. Figure~\ref{Finite} shows
the edge states for a few representative cases. We find that there
exist edge states for both the topologically trivial phase (a, c)
and the topologically non-trivial phase (b, d). The difference between
the two phases is obscured when the magnetic field is not strong enough
for opening a full gap in the phonon spectrum (a, b). In this case,
we can find two counter-propagating modes at each of the edges for
both the phases. However, only one of the modes survives in the gap
regime near $\bm{K}$ ($\bm{K}^{\prime}$) point. The topological
phase transition is accompanied by a change of the propagating direction
of the surviving edge mode. 

The topological difference becomes apparent when we increase the strength
of the magnetic field to open a full gap in the phonon spectrum (c,
d). In this case, the edge modes completely disappear in the gap regime
for the topologically trivial phase, while for the topologically non-trivial
phase, there are two chiral edge modes propagating along the same
direction. The two chiral edge modes are consistent to the Chern number
$C=-2$. One can also clearly see that both the edge modes emerge
from $\bm{\Gamma}$ point, and end near $\bm{K}$ ($\bm{K}^{\prime}$)
point. The topological difference between the two phases lies in the
different ways that the edge modes connect the bulk phonon bands.
For the topological trivial phase, the edges modes connect the same
phonon band, while for the topological non-trivial phase, the edge
modes make inter-band connections. 

We explore the possibility of realizing the TPS in WCs of real materials.
We first check WCs formed in two-dimensional electron gases confined
in semiconductor quantum wells/heterostructure. Some of the hosting
semiconductors, such as AlSb, Al$_{x}$In$_{1-x}$As, InP, and ZnSe,
can be ruled out because their parameters $\alpha_{0}$ and $g$ have
the same sign. On the other hand, other semiconductors such as GaAs,
AlAs, InAs, InSb do have opposite signs for $\alpha_{0}$ and $g$.
Table \ref{tab:pars} shows relevant parameters for them. Unfortunately,
we find that the SOC is too weak for all of these materials. For a
WC stabilized purely by the electron-electron interaction, SOC induced
gaps $\Delta_{\bm{K}}$ are of the order of $10^{-7}$~meV, and the
magnetic field must be weaker than $10^{-7}$~T for a TPS. This is
apparently impossible for real world experimental conditions.

\begin{table}[tb]
\begin{tabular}{|c||c|c|c|c|c|c|}
\hline 
 & $g^{\ast}$  & Ry$^{\ast}$ & $\alpha_{0}/\hbar$  & $\hbar\omega_{0}^{r_{s}=38}$ & $\Delta_{\bm{K}}^{r_{s}=38}$  & $B_{c}^{r_{s}=38}$ \tabularnewline
 &  & (meV) & ($\textrm{eV}^{-1}$) & (meV) & (neV) & ($\mu$T)\tabularnewline
\hline 
\hline 
GaAs & - & $5.882$ & $0.0455$ & $0.0494$ & $0.221$  & $0.127$\tabularnewline
\hline 
AlAs & +  & $20.157$ & $-0.00479$ & $0.169$  & $0.274$  & $0.356$\tabularnewline
\hline 
InAs  & -  & $1.461$ & $0.352$ & $0.0123$  & $0.106$  & $0.0210$\tabularnewline
\hline 
InSb  & -  & $0.590$ & $0.955$ & $0.00495$  & $0.0469$  & $0.00563$\tabularnewline
\hline 
\end{tabular}

\caption{\label{tab:pars}Parameters calculated for a number of semiconductors.
Material parameters are adopted from Ref.~\cite{key-6b}, with $\alpha_{0}\equiv(m^{\ast}/\hbar e)r_{41}^{6c6c}$.
Ry$^{\ast}$ is the effective Rydberg for the material, $\Delta_{\bm{K}}\equiv2\hbar\alpha_{0}\omega_{0}^{2}$
is the gap induced by SOC at $\bm{K}$-point, and $B_{c}$ is the
critical strength of the magnetic field for the topological phase
transition. The values of $\omega_{0}$, $\Delta_{\bm{K}}$ and $B_{c}$
at $r_{s}=38$ are shown, as indicated by the superscripts. The values
of these quantities at other density can be determined by: $\omega_{0}=\omega_{0}^{r_{s}=38}(38/r_{s})^{3/2}$,
$\Delta_{\bm{K}}=\Delta_{\bm{K}}^{r_{s}=38}(38/r_{s})^{3}$, $B_{c}=B_{c}^{r_{s}=38}(38/r_{s})^{3}$.}
\end{table}

We also explore the possibility in hole systems. In this case, the
SOC has a different form due to the band symmetry~\cite{key-6b}.
As a result, the coupling between the momentum and spin is proportional
to $\left[p_{x}(\bm{l})E_{y}(\bm{l})+p_{y}(\bm{l})E_{x}(\bm{l})\right]\sigma$,
instead of $[p_{x}(\bm{l})E_{y}(\bm{l})-p_{y}(\bm{l})E_{x}(\bm{l})]\sigma$
for an electron system. It gives rise to a different $G(\bm{k})=eB/m+\alpha_{0}\left(D_{yy}(\mathbf{\mathbf{k}})-D_{xx}(\mathbf{\mathbf{k}})\right)\sigma$,
in which the SOC contribution vanishes at $\bm{K}$ point. Therefore,
the SOC in a hole system cannot drive a topological phase transition
of phonon.

We also explore the possibility in WCs stabilized by a strong magnetic
field, which quenches the kinetic energy of electrons and favors the
formation of WCs~\cite{Fukuyama1979,Maki1983}. These WCs could be
stabilized in quantum Hall systems at much higher electron densities~\cite{Zhu2010}.
For a typical electron density with $r_{s}\sim1$, we find that (see
Table \ref{tab:pars}) for GaAs, $\Delta_{\bm{K}}\sim12\,\mu\mathrm{eV}$
and $B_{c}\sim7\,\mathrm{mT}$. Other semiconductors have parameters
in similar orders of magnitude. They are still too small to provide
observable physical effects.

All summarized, we conclude that WCs in real semiconductor materials
cannot support a TPS. This is not surprising because as a relativistic
effect, the SOC is always weak. From Table \ref{tab:pars}, we see
that the strength of the SOC must be enhanced at least four orders
of magnitude to reach $\alpha_{0}\omega_{0}\sim1$, a magnitude necessary
for a clear manifestation of the topological effect. This is unfortunately
impossible in real world. 

We argue that one may look for the TPS in WCs with emergent effective
SOC. Actually, SOC is only one of many possible forms of coupling
between orbital motion and internal degrees of freedom. The strengths
of other forms of the coupling are not necessarily constrained by
the relativistic principle, and could be potentially very strong.
An interesting and potentially relevant case could be found in the
fractional quantum Hall systems, in which a new species of WCs, \textit{i.e.},
Wigner crystals of composite fermions, may form~\cite{Archer2013}.
In these systems, the orbital motion of electrons is strongly entangled
with degrees of freedom of all other electrons in the system due to
the strong-correlation nature of the state. One would expect that
the entanglement serves as effective SOC, and gives rise to similar
effects as those predicted in this paper. The effective SOC emerges
from the strong correlation, and its strength is not constrained by
the relativistic principle. This possibility will be left for further
investigations.
\begin{acknowledgments}
This work is supported by National Basic Research Program of China
(973 Program) Grant No. 2015CB921101 and National Science Foundation
of China Grant No. 11325416.
\end{acknowledgments}

\end{document}